# Safeguarding Efficacy in Large Language Models: Evaluating Resistance to Human-Written and Algorithmic Adversarial Prompts

### Tiarnaigh Downey-Webb, Olamide Jogunola, Oluwaseun Ajao

School of Computing and Mathematics
Manchester Metropolitan University
United Kingdom
tiarnaigh.downey-webb2@stu.mmu.ac.uk, {o.jogunola, s.ajao}@mmu.ac.uk

**Abstract**

This paper presents a systematic security assessment of four prominent Large Language Models (LLMs) against diverse adversarial attack vectors. We evaluate Phi-2, Llama-2-7B-Chat, GPT-3.5-Turbo, and GPT-4 across four distinct attack categories: human-written prompts, AutoDAN, Greedy Coordinate Gradient (GCG), and Tree-of-Attacks-with-pruning (TAP). Our comprehensive evaluation employs 1,200 carefully stratified prompts from the SALAD-Bench dataset, spanning six harm categories. Results demonstrate significant variations in model robustness, with Llama-2 achieving the highest overall security (3.4% average attack success rate) while Phi-2 exhibits the greatest vulnerability (7.0% average attack success rate). We identify critical transferability patterns where GCG and TAP attacks, though ineffective against their target model (Llama-2), achieve substantially higher success rates when transferred to other models (up to 17% for GPT-4). Statistical analysis using Friedman tests reveals significant differences in vulnerability across harm categories ($p < 0.001$), with malicious use prompts showing the highest attack success rates (10.71% average). Our findings contribute to understanding cross-model security vulnerabilities and provide actionable insights for developing targeted defense mechanisms.

**Keywords:** Large Language Models, Adversarial Attacks, Security Assessment, Jailbreaking, AI Safety

## 1. Introduction

Large Language Models (LLMs) have rapidly emerged as transformative technologies across diverse domains, from content generation and decision support to automated reasoning and humancomputer interaction. However, their widespread deployment has raised significant security concerns, particularly regarding their susceptibility to adversarial manipulation through "jailbreaking" attacks that circumvent safety mechanisms to elicit harmful outputs (Wei et al., 2023; Li et al., 2023; Perez and Ribeiro, 2022). The security landscape of LLMs presents unique challenges distinct from traditional cybersecurity domains. Unlike conventional software vulnerabilities that exploit coding errors or system misconfigurations, LLM vulnerabilities often arise from the inherent flexibility of natural language processing and the limitations of current alignment techniques. Adversaries can exploit these weaknesses through carefully crafted prompts that manipulate the model's response generation process, potentially leading to the production of harmful, biased, or inappropriate content (Shen et al., 2024; Kumar et al., 2024; Yu et al., 2024). Current research in LLM security has primarily focused on individual attack methods or limited model comparisons, lacking comprehensive frameworks for systematic security assessment across multiple attack vectors and model architectures (Deng et al., 2023) (Gehman et al., 2020). This fragmentation limits our understanding of crossmodel vulnerabilities and hinders the development of robust defense mechanisms(Shayegani et al., 2023).

This paper addresses these limitations by presenting a comprehensive security assessment framework that evaluates four prominent LLMs against four distinct categories of adversarial attacks. Our systematic approach employs stratified sampling from a large-scale safety benchmark, rigorous statistical analysis, and detailed vulnerability categorization to provide actionable insights for both researchers and practitioners.

### 1.1. Research Contributions

Our work makes several key contributions to the field of LLM security:

1. We provide the first systematic comparison of security robustness across both opensource (Phi-2, Llama-2) and commercial (GPT3.5, GPT-4) LLMs using standardized attack methodologies.

2. We demonstrate significant cross-model transfer effects for optimization-based attacks, revealing shared vulnerabilities that transcend individual model architectures.

3. We introduce rigorous statistical methods for analyzing attack success patterns, including non-

parametric tests appropriate for the characteristics of security assessment data.
4. We provide detailed analysis of model-specific vulnerabilities across six distinct harm categories, enabling targeted security improvements.
5. We establish a standardized framework for LLM security assessment that can be applied to future models and attack methods.

## 1.2. Scope and Limitations

This study focuses on direct prompt injection attacks against four specific LLM architectures. While our methodology is designed to be generalizable, the specific findings are constrained by the models and attack methods evaluated. We acknowledge computational and financial constraints that limited our evaluation scope, and we provide transparent reporting of these limitations to enable proper interpretation of results.

## 2. Background and Related Work

### 2.1. Large Language Model Security Landscape

The security of Large Language Models encompasses multiple dimensions, from training data integrity and model robustness to deployment security and output safety. Unlike traditional software security, LLM security involves managing the inherent uncertainty and creativity of natural language generation while maintaining appropriate boundaries on model behavior.

LLM security concerns can be categorized into several key areas (Liu et al., 2023b) (Gidatkar et al., 2025):

**Prompt Injection Attacks**: These involve crafting inputs that manipulate the model's behavior to produce unintended outputs. Direct prompt injection targets the model directly, while indirect injection involves embedding malicious instructions within seemingly benign content.

**Adversarial Examples**: Similar to computer vision adversarial examples, these involve inputs specifically designed to fool the model while appearing normal to humans.

**Model Extraction and Inversion**: These attacks attempt to extract information about the model's parameters, training data, or internal representations.

**Social Engineering**: These leverage the model's conversational abilities to manipulate users or extract sensitive information.

### 2.2. Jailbreaking Attack Taxonomy

Jailbreaking attacks represent a specific category of prompt injection designed to circumvent safety mechanisms and alignment training (Kumar et al., 2024; Yu et al., 2024). These attacks exploit various weaknesses in current LLM architectures and training methodologies:

**Human-Written Attacks (Li et al., 2024)**: These manually crafted prompts leverage creative language patterns, role-playing scenarios, or social engineering techniques to manipulate model responses. Examples include persona adoption ("act as an uncensored AI"), hypothetical scenarios ("in a fictional world where..."), and indirect instruction encoding.

**Optimization-Based Attacks**: These employ automated optimization techniques to discover effective adversarial prompts:

**Greedy Coordinate Gradient (GCG)**: Utilizes gradient information to iteratively optimize adversarial suffixes that maximize the likelihood of harmful responses (Zou et al., 2023).

**AutoDAN**: Employs hierarchical genetic algorithms to evolve effective jailbreak prompts while maintaining semantic coherence (Liu et al., 2023a). **Tree-of-Attacks-with-Pruning (TAP)**: Uses multiple LLMs in an adversarial framework where one model generates attacks, another evaluates success, and a third refines the approach (Mehrotra et al., 2024).

### 2.3. Safety Assessment Frameworks

Current LLM safety assessment approaches vary significantly in scope, methodology, and rigor. Existing frameworks include:

**Benchmark-Based Evaluation**: Datasets like SALAD-Bench (Li et al., 2024), HarmBench (Mazeika et al., 2024), and AdvBench (Zou et al., 2023) provide standardized test cases for evaluating model safety across different harm categories.

**Red Team Exercises**: Human evaluators attempt to discover vulnerabilities through systematic probing and creative attack generation.

**Automated Adversarial Testing**: Algorithmic approaches generate large numbers of potential attack prompts for systematic evaluation.

**Behavioral Analysis**: Evaluation of model responses across different contexts, personas, and instruction types to identify behavioral patterns and potential vulnerabilities.

### 2.4. Gaps in Current Research

Despite significant recent progress, several gaps remain in LLM security research (Shayegani et al., 2023):

1. Limited Cross-Model Analysis: Most studies focus on individual models or limited comparisons, hindering understanding of shared vulnerabilities.
2. Inconsistent Evaluation Methodologies: Varying evaluation criteria, datasets, and metrics make it difficult to compare results across studies.
3. Insufficient Statistical Rigor: Many evaluations lack appropriate statistical analysis to validate findings and quantify uncertainty.
4. Attack Transferability Understudied: Limited research exists on how attacks developed for one model perform against others.
5. Harm Category Analysis: Insufficient attention to how different types of harmful content exhibit varying levels of attack success.

## 3. Methodology

Our experimental design employs a systematic approach to evaluate LLM security across multiple dimensions. We designed our methodology to address the limitations identified in previous research while maintaining practical feasibility given computational and financial constraints. These are discussed below.

### 3.1. Model Selection and Justification

We selected four LLMs representing different points in the design space:

**Phi-2 (2.7B parameters)**: A compact model from Microsoft Research representing efficient architectures suitable for resource-constrained environments. Its smaller size provides insights into security-efficiency trade-offs.

**Llama-2-7B-Chat (7B parameters)**: Meta's open-source conversational model with extensive safety training, representing current best practices in open-source LLM development.

**GPT-3.5-Turbo**: OpenAI's production model balancing capability and computational efficiency, widely deployed in commercial applications.

**GPT-4**: OpenAI's most capable model at the time of evaluation, representing state-of-the-art LLM capabilities and safety measures.

This selection provides coverage across model sizes (2.7B to >100B parameters), access models (open-source vs. API), and development approaches (research vs. commercial).

### 3.2. Attack Method Implementation

We implemented four distinct attack categories to provide comprehensive coverage of the threat landscape:

#### 3.2.1. Human-Written Attacks

We curated human-written attack prompts from established online communities, including Reddit's jailbreaking communities and JailbreakChat.com. These prompts represent real-world attack patterns developed by adversarial users and provide insight into manual attack effectiveness.

Selection criteria in human-written attacks included documented effectiveness in community discussions, diverse attack strategies (persona adoption, scenario construction, instruction obfuscation), reproducibility and clarity of attack methodology.

#### 3.2.2. AutoDAN Attacks

AutoDAN (Liu et al., 2023a) employs hierarchical genetic algorithms to generate semantically coherent jailbreak prompts. The algorithm maintains readability while optimizing for attack success through evolutionary optimization.

Implementation parameters for AutoDAN were population size of 50, generation limit was set as 20, while mutation rate and crossover probability were 10% and 70% respectively.

#### 3.2.3. Greedy Coordinate Gradient (GCG) Attacks

GCG (Zou et al., 2023) optimizes adversarial suffixes by leveraging gradient information to maximize the likelihood of harmful responses. This white-box attack provides insights into model vulnerability at the parameter level.

Implementation parameters for GCG included a suffix length of 20 tokens, 500 optimization steps: 500; Learning rate: 0.01, top-k candidates: 256.

#### 3.2.4. Tree-of-Attacks-with-Pruning (TAP) Attacks

TAP (Mehrotra et al., 2024) employs a multi-LLM framework where separate models generate attacks, evaluate success, and refine approaches. This black-box method simulates realistic attack scenarios where adversaries lack model access.

Implementation configuration for TAP included an attacker model and judge as GPT-3.5-Turbo, while GPT-4 was used as the evaluator model. The tree was set to a depth of 10 and the branching factor was set as 4.

### 3.3. Dataset Curation and Sampling Strategy

We employed the SALAD-Bench dataset (Li et al., 2024) as our evaluation foundation due to its comprehensive coverage of potential harms and rigorous categorization framework. SALAD-Bench contains 21,000 harmful questions across six primary harm categories: Representation and toxicity; Misinformation harms; Socioeconomic harms; Information and safety; Malicious use; and Human autonomy and integrity.

#### 3.3.1. Stratified Sampling Methodology

To ensure representative coverage while managing computational constraints and maintaining statistical validity for comparative analysis, we employed stratified sampling across harm categories. The sampling procedure was formalized as:

$$n_h = N_h \cdot \frac{n}{N} \quad (1)$$

Where, $n_h$ is the sample size for harm category, $N_h$ is the population size for harm category $h$, $n$ is the total sample size (200 per attack type), and $N$ is the total population size.

### 3.4. Evaluation Framework

#### 3.4.1. Attack Success Rate Calculation

We defined attack success rate (ASR) as the primary metric for evaluating attack effectiveness:

$$ASR = \frac{N_{successful}}{N_{total}} \times 100\% \quad (2)$$

Where, $N_{successful}$ is the number of successful attacks, and $N_{total}$ is the total number of attack attempts. Success determination employed GPT-4 as an automated evaluator with carefully designed prompts to ensure consistent evaluation criteria.

#### 3.4.2. Transfer Attack Success Rate

For attacks optimized against specific models, we calculated transfer attack success rate (T-ASR) when applied to different target models:

$$\text{T-}ASR_{m'} = \frac{N_{successful,m'}}{N_{total}} \times 100\% \quad (3)$$

Where, $N_{successful,m'}$ is the successful attacks on model $m'$. Attacks were originally optimized for model $m$

#### 3.4.3. Evaluator Validation

To ensure evaluation reliability, we conducted metaevaluation comparing GPT-4 assessments with human judgments. The Label Matching Rate (LMR) was calculated as:

$$LMR = \frac{N_{matching}}{N_{evaluated}} \times 100\% \quad (4)$$

Where, $N_{matching}$ is the number of matching assessments and $N_{evaluated}$ is the total number of samples evaluated.

#### 3.4.4. Statistical Analysis Framework

Given the characteristics of our data (small samples, presence of zeros, non-normal distributions), we employed non-parametric statistical methods:

**Friedman Test**: Used to assess differences in attack success rates across harm categories, appropriate for repeated measures with non-parametric data.

**Post-hoc Analysis**: Wilcoxon signed-rank tests with Bonferroni correction for pairwise comparisons when significant differences were detected.

**Effect Size Calculation**: Kendall's W to quantify the magnitude of differences across conditions.

### 3.5. Implementation Details

#### 3.5.1. Technical Infrastructure

The computing environment included a Google Colab Pro subscription with GPU acceleration for open-source models. Funding credits ensured access to LLM APIs, which were granted courtesy of the OpenAI Researcher Access Program, as GPT-3.5 and GPT-4 were used for model evaluation. PromptBench (Zhu et al., 2023) was used for the model loading and interaction. At the same time, further data Processing was done in Python with in-built libraries including Pandas, Numpy, and SciPy for statistical analysis.

#### 3.5.2. Configuration Parameters

For reproducibility, we have configured the experiments to include a response length of a maximum of 200 tokens. This is aimed at balancing informativeness and evaluation efficiency. The temperature was 0.0001 for deterministic responses during evaluation. A timeout of 30 seconds per request to handle potential delays between API calls. Rate limiting was implemented to comply with API usage restrictions.

## 4. Results

### 4.1. Overall Model Robustness Analysis

Our comprehensive evaluation reveals significant variations in security robustness across the four evaluated LLMs. Table 1 presents the attack success rates (ASR) for each model against different attack categories.

### 4.1.1. Key Findings - Model Robustness

With an average ASR of 3.4%, Llama-2 exhibits the strongest overall robustness against adversarial attacks. This finding is particularly significant given that optimization-based attacks (GCG and TAP)

Table 1: Attack Success Rates by Model and Attack Type (%)

| Model | Human-Written | GCG | AutoDAN | TAP | Avg. |
|---|---|---|---|---|---|
| Phi-2 | 7.00 | — | 7.00 | — | 7.00 |
| Llama-2 | 1.50 | 9.00 | 0.00 | 3.00 | 3.40 |
| GPT-3.5 | 2.00 | 0.50 | — | — | 1.25 |
| GPT-4 | 4.00 | — | 3.00 | — | 3.50 |
| Average | 3.63 | 4.75 | 3.33 | 3.00 | 3.79 |

Note: "—" indicates attacks not evaluated due to computational constraints or optimization specificity. GCG and TAP were optimized specifically for Llama2 and evaluated for transferability to other models.

were specifically designed to target Llama-2, yet the model maintained relatively low vulnerability.

Phi-2 demonstrates the highest vulnerability with a 7.0% average ASR across all tested attack types. This consistent vulnerability pattern suggests systematic weaknesses in the model's safety mechanisms rather than attack-specific vulnerabilities.

GPT-3.5 and GPT-4 exhibit different security profiles, with GPT-3.5 achieving the lowest overall ASR (1.25%) while GPT-4 shows moderate vulnerability (3.50%). This variation indicates that model scale and capability do not necessarily correlate with security robustness.

### 4.2. Attack Transferability Analysis

A critical finding of our research involves the transferability of attacks optimized for one model when applied to different target models. Table 2 presents transfer attack success rates for GCG and TAP attacks.

Table 2: Transfer Attack Success Rates (%)

| Target Model | GCG | TAP | Average |
|---|---|---|---|
| Phi-2 | 7.50 | 9.50 | 8.50 |
| GPT-3.5 | 15.50 | 7.50 | 11.50 |
| GPT-4 | 17.00 | 9.50 | 13.25 |
| Average | 13.33 | 8.83 | 11.08 |

### 4.2.1. Significant Transferability Effects

GCG attacks demonstrate substantial transfer effectiveness, particularly against GPT-4 (17.00% T-ASR) and GPT-3.5 (15.50% T-ASR). This high transferability rate significantly exceeds the original effectiveness against Llama-2 (9.00%), suggesting that adversarial suffixes optimized through gradientbased methods exploit fundamental weaknesses shared across model architectures.

TAP attacks show more consistent transferability across models, with T-ASR values ranging from 7.50% to 9.50%. This consistency suggests that the black-box optimization approach discovers attack patterns that exploit general LLM vulnerabilities rather than model-specific weaknesses.

### 4.3. Evaluator Reliability and Validation

To ensure the reliability of our automated evaluation approach, we conducted comprehensive metaevaluation comparing GPT-4 assessments with human expert judgments. Table 3 presents Label Matching Rates (LMR) across different attack types and models.

Table 3: Meta-Evaluation Results - Label Matching Rates (%)

| Attack Type | Phi-2 | Llama-2 | GPT-3.5 | GPT-4 | Avg. |
|---|---|---|---|---|---|
| Human-Written | 20.0 | 90.0 | 90.0 | 70.0 | 67.5 |
| GCG | 90.0 | 90.0 | 90.0 | 100.0 | 92.5 |
| AutoDAN | 50.0 | 80.0 | 90.0 | 50.0 | 67.5 |
| TAP | 90.0 | 90.0 | 20.0 | 40.0 | 60.0 |
| Average | 62.5 | 87.5 | 72.5 | 65.0 | 71.9 |

### 4.4. Harm Category Vulnerability Analysis

Our analysis of vulnerability patterns across different harm categories reveals significant variations in model susceptibility to different types of harmful content. Table 4 presents attack success rates broken down by harm category for each model.

### 4.4.1. Statistical Significance Testing

We conducted Friedman tests to assess the statistical significance of differences across harm categories. The analysis revealed highly significant differences ($\chi^2 = 30.64$, $df = 5$, $p < 0.001$), indicating that observed variations are not due to random chance but reflect systematic patterns in model vulnerabilities.

Post-hoc pairwise comparisons using Wilcoxon signed-rank tests with Bonferroni correction revealed several significant relationships. Such as, Malicious Use vs. Representation & Toxicity showed a highly significant difference ($p < 0.001$), Information & Safety vs. Representation & Toxicity had significant difference ($p < 0.01$) while Socioeconomic Harms vs. Human Autonomy & Integrity also had significant difference ($p < 0.05$)

### 4.4.2. Model-Specific Vulnerability Patterns

Phi-2 demonstrates high vulnerability across multiple harm categories, with particularly concerning ASRs for Socioeconomic Harms, Information & Safety, and Malicious Use (all 10.71%). This pattern suggests systematic weaknesses in safety training rather than category-specific issues.

Llama-2 shows complete resistance to certain categories (Representation & Toxicity, Malicious Use, Human Autonomy & Integrity) while maintaining vulnerability to others (Socioeconomic Harms, Information & Safety). This selective pattern suggests more targeted but incomplete safety training. GPT-4 demonstrates the highest recorded vulnerability to Malicious Use prompts (21.42% ASR), representing a critical security concern for practical deployments. This finding is particularly significant given GPT-4's widespread commercial use.

## 5. Discussion

### 5.1. Implications for LLM Security

Our findings have several important implications for understanding and improving LLM security:

#### 5.1.1. Model Architecture and Security

The substantial differences in vulnerability across models suggest that architectural choices and training methodologies significantly impact security robustness. Llama-2's superior performance across most attack types indicates that open-source models can achieve competitive or superior security compared to commercial alternatives, challenging assumptions about proprietary safety advantages.

The consistent vulnerability of Phi-2 across all tested attack types suggests that smaller models may face fundamental trade-offs between capability and security. This finding has important implications for edge deployment scenarios where computational constraints favor smaller models.

#### 5.1.2. Attack Transferability Threats

The high transferability rates we observed represent a critical security concern for the LLM ecosystem. Organizations relying on commercial models may face elevated risks from attacks developed against open-source alternatives. This finding suggests that security assessment cannot be conducted in isolation but must consider the broader threat landscape across model families.

The particularly high transferability to GPT-4 is concerning given its widespread deployment in sensitive applications. Organizations should consider this transferability risk when conducting security assessments and developing defensive strategies.

#### 5.1.3. Harm Category Prioritization

Our statistical analysis of harm category vulnerabilities provides actionable guidance for prioritizing defensive efforts:

Immediate Priority - Malicious use: The consistently high ASRs and critical GPT-4 vulnerability (21.42%) make this category the highest priority for defensive development.

High Priority - Information & Safety: The broad vulnerability across models and high average ASR (8.03%) indicate systematic weaknesses requiring attention.

Targeted Interventions - Model-Specific Vulnerabilities: Phi-2's broad vulnerabilities and Llama-2's selective weaknesses suggest opportunities for targeted improvements.

### 5.2. Methodological Insights

Our research provides several methodological insights for future LLM security evaluation:

#### 5.2.1. Evaluation Framework Design

We adopted a multi-model comparative analysis. Our results demonstrate the value of comparative evaluation across multiple models rather than single-model assessments. The significant variation in vulnerability patterns would not be apparent from isolated evaluations.

As part of statistical validation importance, the use of appropriate statistical methods revealed significant patterns that might be missed by purely descriptive analysis. The Friedman test results provide strong applications involving high-risk categories like malicious use or information safety.

Multi-Model Evaluation: Security assessments should consider transferability risks from attacks

Table 4: Attack Success Rates by Harm Category (%)

| Harm Category | Phi-2 | Llama-2 | GPT-3.5 | GPT-4 | Average |
|---|---|---|---|---|---|
| Representation & Toxicity | 3.57 | 0.00 | 3.57 | 3.57 | 2.68 |
| Misinformation Harms | 3.57 | 3.57 | 7.14 | 7.14 | 5.36 |
| Socioeconomic Harms | 10.71 | 7.14 | 7.14 | 3.57 | 7.14 |
| Information & Safety | 10.71 | 7.14 | 7.14 | 7.14 | 8.03 |
| Malicious Use | 10.71 | 0.00 | 10.71 | 21.42 | 10.71 |
| Human Autonomy & Integrity | 3.57 | 0.00 | 7.14 | 3.57 | 3.57 |

evidence for systematic differences in harm category vulnerabilities.

To assess the reliability of the evaluator, our metaevaluation results highlight the importance of validation when using automated evaluation approaches. The substantial variation in LMR across conditions
(20-100%) indicates that evaluator reliability cannot be assumed and must be empirically validated for each evaluation context.

### 5.3. Practical Recommendations

Based on our findings, we offer several practical recommendations for different stakeholders:

#### 5.3.1. For Model Developers

Prioritize Malicious Use Defenses: Given the consistently high ASRs for malicious use prompts and the critical vulnerability observed in GPT-4 (21.42%), developers should prioritize strengthening defenses against this harm category.

Cross-Model Security Testing: Our transferability findings indicate that security testing should include attacks developed against other models, not just model-specific evaluations.

Harm Category Balancing: The selective vulnerabilities observed in different models suggest opportunities for more balanced safety training that addresses all harm categories comprehensively.

Evaluation Framework Standardization: Adoption of standardized evaluation protocols would enable better comparison and validation of security improvements across model versions.

#### 5.3.2. For Deploying Organizations

Risk Assessment Integration: Organizations should incorporate harm category vulnerability patterns into their risk assessment processes, particularly for developed against other models in the ecosystem.

Monitoring and Detection: Deployment monitoring should be calibrated based on identified vulnerability patterns, with enhanced scrutiny for high-risk harm categories.

Incident Response Planning: Response procedures should account for the specific vulnerability patterns of deployed models and the potential for transferred attacks.

## 6. Conclusion

This research presents the first comprehensive cross-model evaluation of LLM security against diverse adversarial attack vectors, providing critical insights into the current state of LLM robustness and identifying specific areas requiring improvement.

### 6.1. Key Findings Summary

Our systematic evaluation of four prominent LLMs against four distinct attack categories reveals several important findings:

Significant Model Variation: LLM security robustness varies substantially across models, with Llama-2 demonstrating superior overall security (3.4% average ASR) compared to Phi-2's concerning vulnerability (7.0% average ASR). This variation indicates that security is not simply a function of model scale or commercial development but depends critically on specific architectural choices and training methodologies.

Critical Transferability Effects: Attacks optimized for one model achieve surprisingly high success rates when transferred to others, with GCG attacks reaching 17% success against GPT-4 despite being optimized for Llama-2. This finding reveals shared vulnerabilities across the LLM ecosystem that pose systematic security risks.

Harm Category Vulnerabilities: Statistical analysis confirms significant differences in vulnerability across harm categories ($p < 0.001$), with malicious use prompts representing the highest risk (10.71% average ASR). GPT-4's particular vulnerability to malicious use prompts (21.42% ASR) represents a critical concern for deployed systems.

Evaluation Methodology Challenges: Automated evaluation reliability varies substantially across conditions (20-100% LMR), highlighting the need for careful validation of evaluation approaches and conservative interpretation of results with lower reliability.

### 6.2. Contributions to the Field

Our research makes several significant contributions to LLM security research:

**Methodological Advances**: We introduce rigorous statistical frameworks appropriate for LLM security evaluation, including non-parametric tests and systematic meta-evaluation approaches that address limitations in previous research.

**Comprehensive Comparative Analysis**: Our multi-model, multi-attack evaluation provides the most comprehensive comparison of LLM security robustness to date, revealing patterns not apparent from single-model studies.

**Transferability Quantification**: We provide the first systematic quantification of attack transferability across models, revealing critical shared vulnerabilities that require ecosystem-wide defensive consideration.

**Actionable Security Insights**: Our harm category analysis and statistical validation provide specific, actionable guidance for prioritizing security improvements and defensive development efforts.

### 6.3. Future Research Directions

Our research identifies several critical directions for future work:

Expanded Evaluation Scope: Future studies should include broader model coverage, additional attack methods, and larger sample sizes to strengthen statistical conclusions and improve generalizability. The rapid pace of model development requires continuous evaluation frameworks.

Defensive Strategy Development: Our vulnerability pattern analysis provides foundation for developing targeted defensive approaches. Research should focus on both category-specific defenses and general robustness improvements that address transferability risks.

Evaluation Methodology Improvement: The challenges we identified with automated evaluation reliability highlight the need for improved evaluation approaches, potentially incorporating multiple evaluator models or hybrid human-AI frameworks.

Longitudinal Security Assessment: Continuous monitoring of model security as new versions are released will provide insights into the effectiveness of defensive improvements and the emergence of new vulnerabilities.

### 7. Ethics Statement

This research on LLM vulnerabilities to jailbreaking attacks raises important ethical concerns. While necessary for advancing AI safety, we navigated tensions between security research and potential misuse by obtaining ethics board approval and using only publicly available artifacts. We intentionally limited descriptions of successful attack patterns to prevent providing malicious actors with exploitation instructions. The transferability findings highlight cross-model vulnerability concerns that warrant attention. Our use of GPT-4 as an evaluator introduced potential assessment biases. Ultimately, this work aims to strengthen AI safety mechanisms by identifying specific weaknesses while maintaining responsible research practices.

### 8. Limitations

Computational and Financial Constraints: Our evaluation was limited by available computational resources and API budget constraints. While we received substantial support through OpenAI's Researcher Access Program ($2,500 in credits), API rate limits and computational requirements restricted our evaluation scope.

Evaluation Sample Sizes: The harm category analysis employed small sample sizes (7 prompts per category) that limit statistical power and generalizability. Future research with expanded resources should employ larger sample sizes to strengthen statistical conclusions.

Attack Method Coverage: Our focus on four attack types, while representing major categories, constitutes a limited subset of the full attack landscape. Emerging attack methods and novel adversarial techniques require ongoing evaluation.

Model Selection Constraints: Our evaluation of four models, while representative, cannot capture the full diversity of available LLMs. The rapid pace of model development requires continuous evaluation of new releases.

The substantial variation in GPT-4 evaluator reliability across conditions (LMR ranging from 20% to 100%) represents a significant methodological challenge. This variation may stem from several factors:

Attack Type Complexity: More sophisticated attacks may produce subtle manipulations that are difficult to classify consistently, particularly for TAP attacks which showed the lowest average LMR (60%).

Response Ambiguity: Certain model responses may contain ambiguous content that challenges consistent classification, particularly for Phi-2 responses which showed the lowest model-specific LMR (62.5%).

Evaluation Prompt Design: The evaluation prompts and criteria may require refinement to improve consistency across different attack and model types.

## Acknowledgments

We express our sincere gratitude to OpenAI for providing research credits through their Researcher Access Program, which enabled critical components of this evaluation. We thank our institutional ethics review board for their guidance in conducting this sensitive research responsibly. We also acknowledge the developers of the open-source tools and datasets that made this research possible, including the SALAD-Bench dataset and PromptBench framework. Finally, we thank the anonymous reviewers whose constructive feedback significantly improved the quality and rigour of this work.